\documentclass[10pt]{article}
\usepackage{setspace}
\usepackage[margin=1.0in]{geometry}
\usepackage[utf8]{inputenc}
\usepackage{amsmath, amsfonts, amssymb}
\usepackage{graphicx, float}
\usepackage{xcolor}
\usepackage{todonotes}
\usepackage{authblk}
\newcounter{todocounter}
\providecommand{\keywords}[1]{\textbf{Keywords: } #1}

\title{Ergodic Sensitivity Analysis of One-Dimensional Chaotic Maps}
\author[1,3]{Adam A. \'Sliwiak \thanks{Corresponding author, E-mail address: asliwiak@mit.edu}}
\author[2,3]{Nisha Chandramoorthy}
\author[1,3]{Qiqi Wang}
\affil[1]{{\footnotesize Department of Aeronautics and Astronautics, Massachusetts Institute of Technology, Cambridge, MA, 02139, USA}}
\affil[2]{{\footnotesize Department of Mechanical Engineering, Massachusetts Institute of Technology, Cambridge, MA, 02139, USA}}
\affil[3]{{\footnotesize Center for Computational Science and Engineering, Massachusetts Institute of Technology, Cambridge, MA, 02139, USA}}
\date{July 30, 2020}
\begin{document}
%\begin{frontmatter}
\maketitle
\begin{abstract}
\noindent
Sensitivity analysis in chaotic dynamical systems is a challenging task from a computational point of view.  
%Due to high sensitivity to perturbations, a property inherent to chaotic systems, conventional methods cannot be applied to compute the sensitivities of long-time behavior in these systems. More sophisticated techniques, based on dynamical systems theory, have recently been developed for computing the derivatives of long-time averaged quantities, or equivalently, ensemble averages in chaotic systems, to system parameter perturbations.\\ 
In this work, we present a numerical investigation of a novel approach, known as the space-split sensitivity or S3 algorithm. The S3 algorithm is an ergodic-averaging method to differentiate statistics in ergodic, chaotic systems, rigorously based on the theory of hyperbolic dynamics. We illustrate S3 on one-dimensional chaotic maps, revealing its computational advantage over na\"ive finite difference computations of the same statistical response. In addition, we provide an intuitive explanation of the key components of the S3 algorithm, including the density gradient function. 
\end{abstract}


%\msc{37A05, 37D45, 37M25, 49Q12}

%\end{frontmatter}

\section{Introduction}\label{sect:intro}
%[Adam]:
%\item What is sensitivity analysis and why it is useful:
%----
%\nisha{The first sentence must be rephrased. ``Sensitivity analysis is the study of the response of outputs of a certain model to input parameters''.}
%\nisha{May be just start with ``Sensitivity analysis is a discipline that studies the response of outputs of a certain model to changes in input parameters. This involves computing the derivatives of output quantities of interest to specified parameters. ..''}
%----
%Sensitivity analysis is a mathematical tool used to study how sensitive is the output of a certain model with respect to the input parameters in dynamical systems. From the mathematical point of view, its main purpose is to compute the derivative of the output quantity with respect to input parameters.
Sensitivity analysis is a discipline that studies the response of outputs of a certain model to changes in input parameters. It involves computing the derivatives of output quantities of interest with respect to specified parameters.
This mathematical tool is essential in many engineering and scientific applications, as it enables optimal design of structures \cite{barbosa-example-structures, zhang-shapeopt} and fluid-thermal systems \cite{balagangadhar-example-fluids}, analysis of heterogeneous flows \cite{shiqing-flows}, supply chain management \cite{wang-example-suppluchains}, estimate errors and uncertainties in measurement, modeling and numerical computations \cite{benke-example-error,wang-phdthesis}.
%----
%\nisha{Cite some paper that is related to sensitivity analysis, from the journal we are submitting to. The list of references should include a few from this particular journal and the easiest place to add them would be here.}
%----
%\item When do we need to perform sensitivity analysis of statistics in chaotic systems:
%\nisha{Examples of sensitivity analysis in non-chaotic systems are probably not relevant.}

For example, consider an optimal design problem in structural mechanics involving an elastic truss under loading. In such problems, the ultimate goal might be to approximate derivatives of constrained functions (e.g. resulting displacements) with respect to design parameters (e.g. bar cross-sections). Such derivatives can be approximated using analytical methods, as well as finite differences \cite{kirsch-structures}. 
%For example, consider a pre-notched steel plate subjected to tensile forces. In this particular case, we might be interested how the J-integral is sensitive to the imposed forces. Combining sensitivity analysis and classical finite element simulation, one can provide critical insights into relation of the crack length and external forces \cite{taroco-fracture}. 
A classic example from fluid mechanics is a turbulent flow past a rigid object, in which the sensitivity of the drag (resistance) forces with respect to the Reynolds number and other flow parameters \cite{blonigan-phdthesis, ni-jfm}, is of interest. Particularly, aerospace engineers use the computed sensitivity in the design of airfoils \cite{geng-aero}. Both mechanical phenomena described above are governed by strongly nonlinear dynamical systems, however the latter features an extra difficulty, namely the chaotic behavior.      

%\item Introduce challenges when performing sensitivity analysis of statistics in chaotic systems:
%\nisha{No need of the prelude ``extremely difficult'', jump directly to the why}
Computing such sensitivities in chaotic dynamical systems is a challenging task. The primary issue is the so-called {\it butterfly effect}, which is a large sensitivity of the system to initial conditions. This concept is associated with the classical study of Edward Lorenz on climate prediction \cite{lorenz-climate}.
%\nisha{I would rephrase the previous sentence: } 
Quantitatively, it means that any two points initially separated by an infinitesimal distance diverge at an exponential rate. This implies the prediction of far-future states in chaotic phenomena is hardly possible. We observe this phenomenon in daily weather forecasts, as the predictions of several weeks forward tend to be highly inaccurate.  However, we are sometimes interested in predicting the response of long-time averaged behavior, to perturbations \cite{blonigan-phdthesis,ni-jfm, chandramoorthy-turb}. 
%Due to the chaotic behavior, the quantity of interest is usually the long-time average of a certain physical output, rather then its instantaneous values. 
%---Don't mention differentiability
%Moreover, it has been shown that long-time averages are not necessary differentiable, even in simple one-dimensional chaotic systems \cite{blonigan-pdf}. 

%\item Introduce existing literature, ensemble sensitivity, fluctuation dissipation,
%shadowing method.  Discuss their respective weaknesses.
In the last few decades, there have been different attempts to compute  sensitivities of long-term averages in chaotic systems. The conventional methods \cite{jameson-conventional,cao-conventional}, which require solving either tangent or adjoint equations, fail if the time-averaging window is large. Due to {\it butterfly effect},
%\nisha{almost }
almost every 
%\nisha{infinitesimal} 
infinitesimal perturbation to the system expands exponentially, and therefore the sensitivity computed using the tangent or adjoint solutions grows equally fast. A more successful family of methods utilize the concept of shadowing trajectories. Methods like least squares shadowing (LSS) \cite{wang-lssoriginal, wang-lssconvergence} or its computationally cheaper variant, known as non-intrusive least squares shadowing (NILSS) \cite{ni-nilss}, provide accurate derivatives of long-time averages in many small- and large-scale problems, e.g. weakly turbulent flows \cite{ni-jfm}. However, shadowing methods have been proven to have a systematic error, which can be non-zero 
%\nisha{not large, but non-zero.} 
if the connecting map
%\nisha{diffeomorphism $\to$ the connecting map between the .. }
between the base and shadowing trajectory is not differentiable
%\nisha{differentiable, and this is typically the case}
\cite{ni-systematicerror}. Some approaches adopt the Fluctuation-Dissipation Theorem, which is widely used in the statistical equilibrium analysis of turbulence, Brownian motion, and other areas \cite{kubo-fluctuation}. Unfortunately, they are inexact as well, when they do not assume specific properties of the physical systems, e.g. Gaussian distribution of the equilibrium state \cite{abramov-fluctuation}. Moreover, they require solving costly Fokker-Planck equations, which makes them infeasible for large systems \cite{blonigan-pdf}. Another group of methods for sensitivity analysis are trajectory-based and utilize Ruelle's linear response formula \cite{ruelle-original}. Many of these techniques, generically referred to as ensemble methods, solve tangent/adjoint equations and compute ensemble average over a trajectory to estimate the sensitivity \cite{eyink-ensemble,lea-ensemble}. The two major drawbacks of ensemble-based methods is that they exhibit slow convergence since they 
%\nisha{and $\to$ since they} 
suffer from exponentially increasing variance of the tangent/adjoint equations \cite{eyink-ensemble,chandramoorthy-turb}.   

%\item Introduce the S3 method in a few sentences.
Space-split sensitivty (S3) is an alternative trajectory-based method that uses Ruelle's formula \cite{chandramoorthy-s3}. However, unlike the ensemble methods, it does not manifest the problem of unbounded variances. Moreover, the S3 method does not assume that the probability distribution in state space is of a particular type (e.g. Gaussian), and also does not rely on directly estimating the probability distribution by e.g. discretizing phase space.
%Moreover, the performance of the S3 method does not depend on the properties of a system, as the sensitivity is computed through sampling along the trajectory.
%\nisha{Moreover, the S3 method does not assume that the probability distribution in state space is Gaussian, and also does not rely on directly estimating the probability distribution in phase space} 
 In the paper, we will closely review the basic concepts of the S3 method in the context of one-dimensional chaotic maps.   

%\item Outline this paper.
%\nisha{use labels}
In Section \ref{sect:maps} of this paper, we review two representative one-dimensional maps that exhibit chaotic behavior, namely the sawtooth and cusp map. The space-split sensitivity method is derived in Section \ref{sect:s3}. Section \ref{sect:g} focuses on the interpretation and computational aspects of the density gradient, which is a key quantity appearing in the S3 method. Section \ref{sect:sens} demonstrates numerical examples showing sensitivities generated using the S3 method. Finally, Section \ref{sect:concl} concludes this paper.

%-----------------Section 2 ----------------------------
\section{Parameterized one-dimensional chaotic maps and their statistical dependence on parameter}\label{sect:maps} 
%[Adam]:
%\item Outline the objectives of this section -- to establish test cases and reference
%truth for sensitivity analysis:
\noindent
%----
%\nisha{In this section, we introduce two families of perturbed one-dimensional chaotic systems. We review the concepts of Lyapunov exponents and ergodicity 
%through numerical illustrations on the two maps.}
%----
%The purpose of this section is to review the simplest, one-dimensional, discrete chaotic systems in the form
In this section, we introduce two families of perturbed one-dimensional chaotic systems that can generally be expressed as
\begin{equation}\label{2:1Dmap}
x_{k+1}=\varphi(x_{k};s),\hspace{2cm}x_{0} = x_{\rm init},
\end{equation}
where $x_{\rm init}$ is a given initial condition, while $s$ denotes a scalar parameter. Let $J$ be a scalar observable. Our quantity of interest is the infinite-time average or \emph{ergodic} average of $J$,
\begin{equation}\label{2:longtime}
    \langle J\rangle := \lim_{N\to\infty}\frac{1}{N}\sum_{i=0}^{N-1}J(x_{i};s).
\end{equation}
In particular, we focus on the relationship between $\langle J \rangle$
and the parameter $s$ for $\varphi$. In addition, we review the concepts of Lyapunov exponents and ergodicity 
through numerical illustrations on the two maps.
%\item Introduce the saw-tooth map and its parameterization. What makes it chaotic? Introduce Lyapunov exponents and its computation.
%\nisha{We analyze the Lyapunov exponents for a family of sawtooth and cusp maps, and describe their statistical behavior.}

\subsection{Perturbations of the sawtooth map and their Lyapunov exponents}
\begin{figure}[H]
    \centering
    \includegraphics[width=0.6\textwidth]{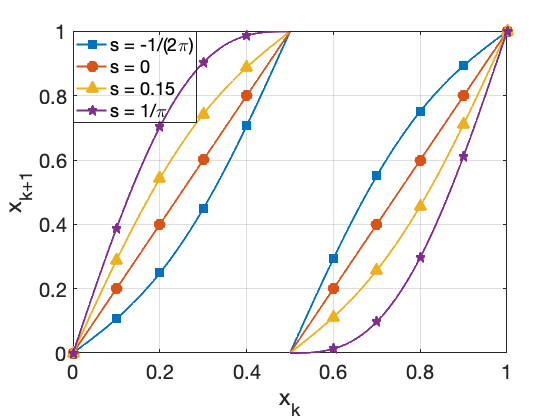}
    \caption{The sawtooth map at different values of parameter $s$. Note if $s=0$, we obtain the classical Bernoulli shift.}
    \label{fig:sawtooth_map}
\end{figure}

We consider as our first example, perturbations of the sawtooth map, also known as the dyadic transformation, defined in the following way:
\begin{equation}\label{2:sawtooth}
x_{k+1}=\varphi(x_{k};s) =
2x_{k}+s\hspace{1mm}\mathrm{sin}(2\pi x_{k}) \; {\rm mod}\; 1,
\;\;\;\; x_{k}\in [0,1).
\end{equation}
It is a periodic map that maps $[0,1)$ to itself. Figure \ref{fig:sawtooth_map} illustrates the sawtooth map for different values of the parameter $s$. 
%----
%\nisha{Roughly speaking, a \emph{chaotic} map shows high sensitivity to initial conditions. 
%For example, consider $s=0,$ and two phase points 
%$x$ and $x + \delta x$. Under one iteration of 
%the map, these two points are now separated by a distance 
%of $2 \delta x.$ Thus, in the limit $\delta x \to 0,$
%a trajectory that is infinitesimally separated from $x$ 
%at $n=0$ moves away from the trajectory of $x$ at an 
%exponential rate, $\log 2.$ This exponential growth 
%of perturbations to the state is the signature of chaotic 
%systems and is measured by the rate of asymptotic growth, 
%known as LE. More rigorously, LEs are defined by : Eq.4.
%which clearly indicates that the infinite-time averaged 
%rate of growth converges to a constant. 
%We say that a map is chaotic when its LE is positive. 
%As mentioned the sawtooth map, introduced in Eq. 3, using this 
%definition, is chaotic, with a LE of $\log 2$, when s=0.
%}
%----

\begin{figure}[H]
    \centering
    \includegraphics[width=0.6\textwidth]{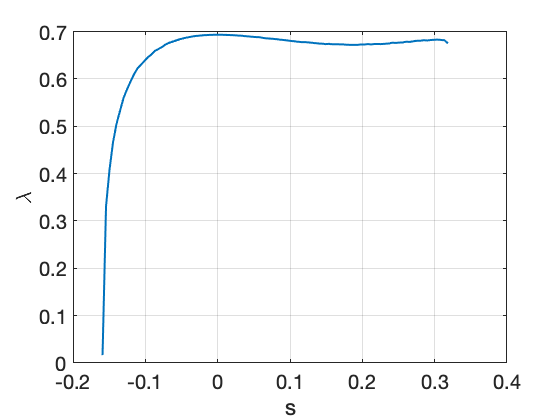}
    \caption{Relation between the Lyapunov exponent $\lambda$ and parameter $s\in \left[-\frac{1}{2\pi},\frac{1}{\pi}\right]$ for the sawtooth map.}
    \label{fig:sawtooth_le}
\end{figure}

A natural question that arises is whether the chosen is map actually chaotic. 
%A rigorous answer to that question can be provided through the analysis of the Lyapunov exponent, usually denoted as $\lambda$, which is a key quantity in the analysis of dynamical systems. 
Roughly speaking, a \emph{chaotic} map shows high sensitivity to initial conditions. 
For example, consider $s=0,$ and two phase points 
$x$ and $x + \delta x$. Under one iteration of 
the map, these two points are now separated by a distance 
of $2 \delta x.$ Thus, in the limit $\delta x \to 0,$
a trajectory that is infinitesimally separated from $x$ 
at $n=0$ moves away from the trajectory of $x$ at an 
exponential rate of $\log 2\approx 0.693.$ This exponential growth 
of perturbations to the state is the signature of chaotic 
systems and is measured by the rate of asymptotic growth, 
known as the Lyapunov exponent (LE) and denoted by $\lambda$. 
More rigorously, the Lyapunov exponent is defined by
\begin{equation}
\label{2:LE}
\lambda(s) = \lim_{n\to\infty}\frac{1}{n}\sum_{k=0}^{n-1}\log\left|\frac{\partial \varphi}{\partial x}(x_{k};s)\right|,
\end{equation}
which clearly indicates that the infinite-time averaged 
rate of growth converges to a constant. 
We say that a map is chaotic when its LE is positive. 
%As mentioned the sawtooth map, introduced in Eq. \ref{2:sawtooth}, using this 
%definition, is chaotic, with a LE of $\log 2$, when s=0.
%The Lyapunov exponent measures the sensitivity of the system to perturbations of its initial conditions. Therefore, it can be used to verify whether a system is chaotic or not. 
%To illustrate its meaning, consider two trajectories of the sawtooth map (Eq. \ref{2:1Dmap}), whose states at step $k$ are seperated by an infinitesimal distance $\delta x$. Given $\lambda$, we can estimate that at step $k+1$ this distance rescales by the factor of $e^{\lambda}$. In other words, the Lyapunov exponent determines how fast two infinitesimally close trajectories separate. Therefore if $\lambda >0$, then the system is chaotic. 
%An efficient numerical method for computing $\lambda$ has been proposed in \cite{benettin-le}. In this paper, however, we consider only one-dimensional maps and therefore we can directly apply the following formula:
%\begin{equation}
%\lambda(s) = \lim_{n\to\infty}\frac{1}{n}\sum_{i=0}^{n-1}\log\left|\frac{\partial %\varphi}{\partial x}(x_{n};s)\right|,
%\end{equation}
Formula \ref{2:LE} requires computing the derivative of the map at points along a trajectory. Note that the value of the Lyapunov exponent does not depend on initial condition $x_{\rm init}$, nor on the step $k$. 
%----
%\nisha{use n/k consistently}
%----
Figure \ref{fig:sawtooth_le} shows that $\lambda>0$ for all $s\in(-\frac{1}{2\pi}, \frac{1}{\pi}]$ meaning that the sawtooth map is chaotic in this regime. This can be easily justified by the observation that $\frac{\partial\varphi}{\partial x}\geq 1$ for all $x\in[0,1]$, when s is in this regime. 
%----
%\nisha{when s is in this regime}. 
%----
%One can also observe this relation is smooth, which is not necessarily the case for general one-dimensional chaotic maps.
%----
%\nisha{get a reference - may not be true.}
%----

%----
%\nisha{Note that at $s=0$, the sawtooth map always appears to converge to a fixed point, after some iterations, when simulated numerically. This is because all machine-representable numbers with a fixed number of digits after the binary point, are dyadic (binary-rational) numbers, which converge to a fixed point under this map, since the sawtooth map is simply a leftshift operation at $s=0$. However, these are not typical trajectories of the Sawtooth map. In order to simulate a typical chaotic trajectory, refer to the appendix.}
%----

In case of the classical Bernoulli shift, i.e. when $s=0$, repetitive the sawtooth map always appears to converge to a fixed point, after some iterations, when simulated numerically. This is because all machine-representable numbers with a fixed number of digits after the binary point, are dyadic-rational numbers, which converge to the fixed point 0, under this map, because the sawtooth map at $s=0$ is simply a leftshift operation on binary digits. More details about this problem and possible remedies can be found in Appendix \ref{appendix:floatingpoint}. Note also that if $s=0$ and $x_{\rm init}$ is rational, the forward orbit of $x_{\rm init}$ would either converge to a fixed point or be periodic, containing a finite number of distinct values within the interval $[0,1)$. For example, if $x_{\rm init}=0.1$, then all future states belong to a four-element set, $\{0.2, 0.4, 0.6, 0.8\}$, and $x_{k}=x_{k+4}$ for all $k>0$. This is an example of an unstable periodic orbit; in this paper, we are interested in chaotic orbits, which are aperiodic and unstable to perturbations.
%----
%\nisha{This is an example of an unstable periodic orbit; in this paper, we are interested in chaotic orbits, which are aperiodic and unstable to perturbations.}
%----
%Although this is an instance of an unstable system, the analysis of periodic orbits is beyond the scope of this paper.
%----------------------------------
% TO BE DELETED
 %An efficient algorithm to compute Lyapunov exponents in an {\it arbitrary} chaotic system have been %proposed in [CITE Benettin's paper]. In case of one-dimensional maps, this procedure only requires %solving \ref{2:1Dmap} and its tangent equations,
%\begin{equation}\label{2:1Dmap_tangent}
%v_{k+1}=\frac{\partial\varphi(x_{k};s)}{\partial x_{k}}v_{k},\hspace{2cm}v_{0} = v_{init},
%\end{equation}
%where $x_{init}$ and $v_{init}$ are {\it arbitrary} initial conditions. The exponent can be approximated %using
%\begin{equation}
%\lambda \approx \frac{1}{N}\sum_{i = 0}^{N-1}\log|v_{i}|.
%\end{equation}
%It can be shown \ref{2:sawtooth} has a non-positive $\lambda$ and therefore is non-chaotic if $s \leq %-\frac{1}{2\pi}$. Inside the interval $s\in[-\frac{1}{2\pi}, \frac{1}{pi}]$ $\lambda$ is always %positive meaning the system is chaotic and, moreover, there is a smooth dependence of $\lambda$ and $s$ %(see Figure \ref{fig:sawtooth_le}).
%---------------------------------

%\item Introduce the cusp map and logistic map and their parameterization. 
\noindent
%\nisha{A more standard name for the cusp map?} Adam: "cusp" is a standard name for maps of this type
\subsection{A family of Cusp maps and their Lyapunov exponents}
\begin{figure}[H]
    \centering
    \includegraphics[width=0.6\textwidth]{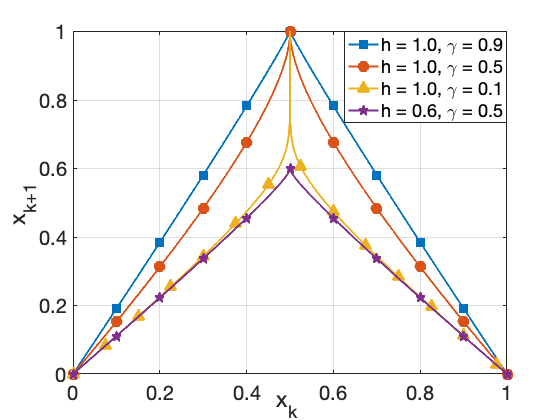}
    \caption{The cusp map at different values of parameters $h$ and $\gamma$. Note all the curves include points $(0,0)$ and $(1,0)$, while the tip is located at $(0.5,h)$. If $h=\gamma=1$, the map is piecewise linear, and this particular case is usually referred to as the tent map.}
    \label{fig:cusp_map}
\end{figure}

\begin{figure}[H]
    \centering
    \includegraphics[width=0.6\textwidth]{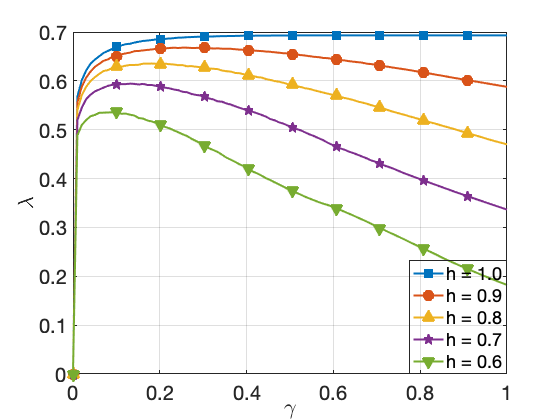}
    \caption{Relation between the Lyapunov exponent $\lambda$ and parameter $\gamma\in [0,1]$ for the cusp map.}
    \label{fig:cusp_map_le_1D}
\end{figure}

Another example of a chaotic map is the cusp map $\varphi:\hspace{1mm}[0,1]\to[0,h]$ defined as follows,
\begin{equation}
\label{eqn:cusp}
    x_{k+1} = \varphi(x_{k};h,\gamma)
  = h - \left|\frac12 - x_{k}\right|
  - \left(h-\frac12\right)
    \left|1 - 2 x_{k}\right|^{\gamma}.
\end{equation}
%----
%\nisha{The cusp map is a two-parameter map with $s = [h, \gamma]$, where $h$ is the height ...}
%----
The above function produces a spade-shaped graph, as shown in Figure \ref{fig:cusp_map}. The cusp map is a two-parameter map with $s = \{h, \gamma\}$, where $h$ is the height, while $\gamma$ is a parameter that determines the sharpness of the tip. We use the definition Eq. 4 to compute the LE of the cusp map at different values of $h$ and $\gamma$. From the positivity of the Lyapunov exponent shown in Figure \ref{fig:cusp_map_le_1D}, we see that the cusp map is always chaotic if $\gamma\in[0,1]$ and $h\geq0.6$.
%[Adam: can you plot the map for several heights $h$ and exponents $\gamma$
%to show the readers what the map looks like and how it depends on the
%two parameters?]
%----
%\nisha{piecewise, not hyphenated, in the caption in Figure 1.}
%----

%----
%\nisha{We use the definition Eq. 4 to compute the LE of the cusp map at different values of $h$ and $\gamma$. From Figure, we 
%see that the cusp map is always chaotic if...}
%----

%Note also the relation $\lambda$ vs. $\gamma$ is smooth, therefore we will focus on this particular parameter domain in our further analysis. 
%----
%\nisha{The smoothness of $\lambda$ vs $\gamma$ is not relevant to sensitivity analysis. please refer to the theoretical literature on the result on smoothness of LEs, in uniformly hyperbolic systems.}
%----

%----
%\nisha{The cusp map is a 1-dimensional representation of the 
%3-dimensional Lorenz'63 system, that shows the latter\'s essential features. Specifically, the iterates of the cusp map are local maxima of the third coordinate of the Lorenz'63 system.}
%----
Historically, the 
cusp map has been used as a one-dimensional representation of the three-dimensional Lorenz'63 system \cite{lorenz-climate}, a set of ordinary differential equations used as a model for atmospheric convection. Specifically, the iterates of the cusp map are local maxima of the third coordinate of the Lorenz'63 system \cite{mehta-cusp}. 

\subsection{Ergodic probability distributions}
\begin{figure}[H]
    \centering
    \includegraphics[width=0.6\textwidth]{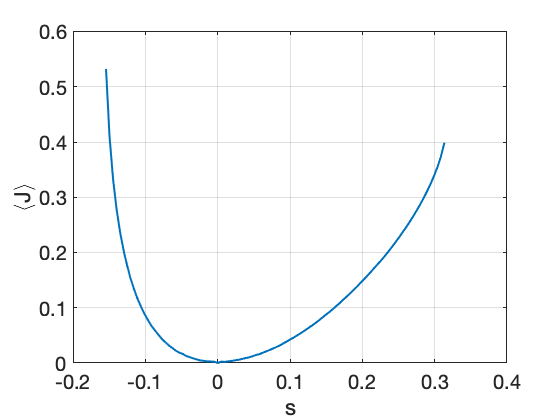}
    \caption{Long-time averaged behavior with respect to the map parameter for the sawtooth map. The objective function itself does not depend to $s$, and is defined as $J(x)=\cos(2\pi x)$. In our computations, $J$ is averaged over 100 million samples.}
    \label{fig:sawtooth_jav}
\end{figure}

\begin{figure}[H]
    \centering
    \includegraphics[width=0.6\textwidth]{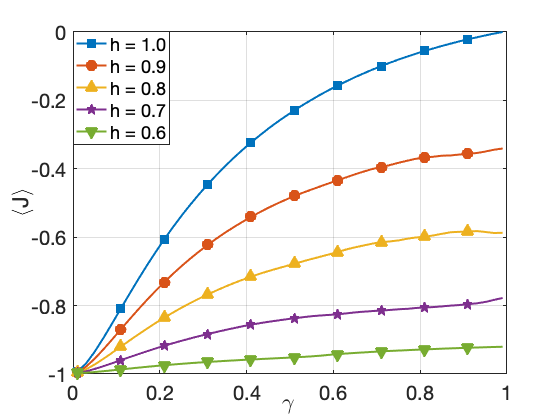}
    \caption{Long-time averaged behavior with respect to the map parameters for the cusp map. The objective function itself does not depend to $h$ nor $\gamma$ and is the same as in the previous example, i.e. $J(x)=\cos(2\pi x)$.}
    \label{fig:cusp_jav}
\end{figure}

The long-time average of the objective function, $\langle J\rangle$, was calculated using 100 million iterates of the map, 
with the initial condition chosen uniformly, at random 
between (0,1), in the following way:
\begin{equation}\label{2:longtime_num}
    \langle J\rangle \approx \frac{1}{N}\sum_{i=0}^{N-1} J(x_{i}),
\end{equation}
where $x_{i+1}=\varphi(x_{i})$. We choose a sufficiently large $N$ to 
ensure that the right hand side converges to a fixed value, within 
numerical precision. Figures \ref{fig:sawtooth_jav} and \ref{fig:cusp_jav} illustrate examples of the mean statistics (i.e. long-time averages) and their dependence on the map parameters for the sawtooth and cusp map, respectively.  

In the computation of the long-time averages of the objective function, we used the concept of ergodicity. This property guarantees that long-time averages do not depend on the initial conditions. That is, the time average of the objective function (right hand side of Eq. \ref{2:longtime_num}) converges, as $N\to \infty$, to a value independent of the initial condition $x_0$, for almost every $x_0$ chosen uniformly between $(0,1)$. This limit equals the expected value of the same objective function over an ensemble of initial conditions distributed according to 
an \emph{ergodic}, \emph{invariant} probability distribution $\rho$. This probability distribution $\rho$ is invariant under $\varphi, $ in the sense that for any 
open interval $A \subset (0,1)$, $\rho(A) = \rho(\varphi^{-1}(A)).$ 
In addition, $\rho$ is defined by the fact that expectations with respect to $\rho$ are the same as infinite-time averages starting from a point uniformly distributed in the unit interval. Such a probability distribution
$\rho,$ is known as the SRB distribution \cite{young-srb}, and only sometimes coincides with the uniform distribution (it does e.g. for the sawtooth map at $s=0$).
%----
%\nisha{, $ in the sense that for any 
%open interval $A \in (0,1)$, $\mu(A) = \mu(\varphi^{-1}(A)).$ 
%In addition, it is defined by the fact that expectations with respect to $\mu$ are the same as infinite-time averages starting from a point uniformly distributed between (0,1). Such a measure 
%$\mu,$ is known as the SRB measure, and only sometimes coincides with the uniform distribution.} 
%----
\noindent
The above description can be mathematically rephrased as follows, for almost 
every $x_0$ uniformly distributed in $(0,1)$,
\begin{equation}\label{ergodic}
    \langle J\rangle = \lim_{N\to\infty}\frac{1}{N}\sum_{i=0}^{N-1}J(x_{i}) = \int_{U}J(x)\,\rho(x)\;dx.
\end{equation}
Thus, in ergodic systems, there exist two alternative ways of computing the long-time average, either through the averaging of the time-series or ensemble averaging. The latter requires prior computation of the probability distribution, which will be explained and illustrated in the next sections. Using these preliminary concepts, we will review the space-split sensitivity method to compute the derivative of $\langle J\rangle$ with respect to the map parameter.
%----
%\nisha{General comment: if you take $x$ Lebesgue-a.e, the infinite time average of continuous functions is equal to 
%the ensemble average wrt to the SRB measure}
%----

%\end{itemize}

%----------------End of Section 2-----------------------

%---------------------- Section 3-----------------------
\section{Split Space Sensitivity analysis of one-dimensional maps}\label{sect:s3} 
%[Adam]:
%\begin{itemize}
%\item Explain Ruelle's linear response equation, specifically for a 1D maps
%$\varphi:U\to U$ parameterized by $s$, with an quantity of interest
%$J:U\to R$.
In \cite{ruelle-original}, Ruelle rigorously derived a formula for the derivative of the quantity of interest, $\langle J\rangle$, with respect to the map parameter $s$. This expression is an ensemble average (or expectation) with respect to $\rho$, which can be simplified for one-dimensional maps $\varphi:U\to U$ to
\begin{equation} \label{ruelle} 
    \frac{d}{ds} \int_U J(x) \,\rho(x)\;dx
  = \sum_{k=0}^{\infty} \int_U
    f(x)\,
    \frac{d \big( J\circ\varphi_{k}\big)(x)}{dx} \,\rho(x)\;dx
\end{equation}
where
\begin{equation} \label{xspace}
    f(x) := \frac{\partial\varphi\big(\varphi^{-1}(x)\big)}{\partial s}
\end{equation}
reflects the parameter perturbation of the map, while $U$ refers to the unit interval $[0,1)$.
%----
%\nisha{Here, $U$ refers to the unit interval $[0,1).$. SRB measure for a manifold without boundaries.}
%----
%----
%\nisha{A direction evaluation of Eq. 8 is computationally cumbersome for the following reason... }
%----
%Using tangent/adjoint solvers, one could directly implement Eq. \ref{ruelle}--\ref{xspace} to find desired sensitivities. This approach, however, would pose some computational difficulties. 
A direct evaluation of Eq. 8 is computationally cumbersome for the following reason. Notice that the integrand of the right hand side of Eq. \ref{ruelle} involves a derivative of the composite function that can be expanded using the chain rule to the form
\begin{equation}\label{ruelle_product}
    \frac{d (J\circ \varphi_{k})(x)}{dx} = \Big( \frac{dJ}{dx} (\varphi_k (x))\Big) \prod_{j=0}^{k-1}\frac{\partial\varphi}{\partial x}(\varphi_j(x)).
\end{equation}

As discussed earlier, for a large $k$, the product of the derivatives exponentially grows with $k$. However, Ruelle's series converges due 
to cancellations of these large quantities, upon taking an ensemble average. 
This problem makes the direct evaluation of Ruelle's formulation computationally impractical since a large number of trajectories are needed for these cancellations. More precisely, since for a large $k$, 
\begin{equation}
\frac{d \big( J\circ \varphi_k \big)}{dx}(x) \sim {\cal O}(e^{\lambda k}), 
\end{equation}
at almost every $x$, we need to increase the number of trajectories by a factor of ${\cal O}(e^{2 \lambda k})$ in order 
to reduce the mean-squared error in a linear fashion.
%----
%\nisha{This problem makes the direct evaluation of Ruelle's formulation computationally impractical since a large number of trajectories are needed for these cancellations. More precisely, since for a large $k$, 
%$$\frac{d J\big(\varphi^k \big)}{dx}(x) \sim {\cal O}(e^{\lambda k})$$, at almost every $x$, we need to increase the number of trajectories by a factor of ${\cal O}(e^{2 \lambda k})$ in order 
%to reduce the mean-squared error in the ensemble average in a linear fashion.}
%----
%Although the series in Eq. \ref{ruelle} has been proven to converge due to cancellations over a long trajectory as $k\to\infty$,
%this problem makes classical Ruelle's formulation computationally intractable.
For example, consider the sawtooth map with $s\in[-\frac{1}{2\pi},\frac{1}{\pi}]$. In this case, $(\partial\varphi/\partial x)\in[1,4]$. One can easily verify that even for moderate values of $k,$ an overflow error is encountered. Another challenge is that the evaluation of the SRB distribution requires expensive computation of map probability densities \cite{blonigan-pdf}.
%----
%\nisha{evaluation of the SRB measure}
%----
In a recent study \cite{chandramoorthy-s3}, Ruelle's formula has been reformulated to a different ensemble average, known as the S3 formula. There, the latter formula has been derived for maps of arbitrary dimension, and is based on splitting the total sensitivity into that due to stable and unstable perturbations.
%----
%\nisha{the one dimensional perturbation, is by defintion, unstable}.
%----
Note the notion of splitting the perturbation space is irrelevant for 1D maps, and the one-dimensional perturbation is, by defintion of chaos, unstable. Therefore we will skip some aspects of the original derivation, and 
note that our derivation represents only the unstable component
of sensitivity in \cite{chandramoorthy-s3} specialized to 1D.
%----
%\nisha{and 
%note that our derivation represents only the unstable component
%of sensitivity in \cite{chandramoorthy-s3} specialized to 1D.}
%----

%\item Re-derive S3 formula in the specific case of one-dimensional chaotic map,
%i.e., not
%considering stable manifold, no need for decomposition into CLV directions,
%easy integration-by-parts.  Derive the equation:
The S3 formula, corresponding to equations \ref{ruelle}--\ref{xspace}, can be expressed as follows:
\begin{align} \label{s3eqn}
    \frac{d}{ds} \int_U J(x) \,\rho(x)\;dx
  &= -\sum_{k=0}^{\infty} \int_U 
  \nabla_{\rho} f(x) \, J\big(x_{k})\; \rho(x)\;dx ,
\end{align}
where 
\begin{equation} \label{rho_div}
    \nabla_{\rho} f(x) := \frac{1}{\rho(x)}
    \frac{d\big(\rho(x)\,f(x)\big)}{dx}
  = \frac{df}{dx}(x) + f(x)\, g(x),
\end{equation}
%----Don't need it anymore
%where $\rho(x)$ is the probability density function of the SRB distribution
%\begin{equation}
%    \rho(x) := \frac{d\mu(x)}{dx}
%\end{equation}
%----
and,
\begin{equation} \label{rhogrelation}
    g(x) := \frac{1}{\rho(x)} \, \frac{d\rho}{dx}(x).
\end{equation}

For one-dimensional maps, the derivation is simple, as it requires integrating \ref{ruelle} by parts and the fact that the integral of $df/dx$ at the boundary of $U$ vanishes; see Appendix \ref{appendix:s3rederivation} for the full derivation. We observe that both $J$ and $df/dx$ have their analytical forms.  However, the function $g(x)$, which will be referred to as \emph{density gradient}, does not have a closed-form expression, since the SRB distribution 
$\rho$, is unknown. The density gradient $g$ represents the variation in phase space, of the logarithm of $\rho(x)$,
\begin{equation}\label{eqn:iterativeg}
    g(x) := \frac{1}{\rho(x)} \, \frac{d\rho}{dx}(x)
          = \frac{d \log\rho(x)}{dx}.
\end{equation}
In the next section, we focus on further interpreting $g(x)$, its computation and verification on the 1D maps introduced in Section \ref{sect:maps}.

%----------------End of Section 3-----------------------

%---------------------- Section 4-----------------------

\section{Computation of density gradient}\label{sect:g} 
%[Adam]:
%\begin{itemize}
%Re-derive the recursive equation for $g(x)$ specifically for 1D case:
In this section, we focus on the density gradient function, denoted by $g(x)$. First, we present a computable, iterative scheme for $g(x)$. Moreover, we provide an intuitive explanation for $g$ and visualize it  on the maps introduced in Section \ref{sect:maps}.

Based on the S3 formula (Eq. \ref{s3eqn}), we can conclude the following recursive relation
\begin{equation}\label{iterative_g}
    g\big(\varphi(x)\big) = \frac{g(x)}{d\varphi(x)/dx}
  - \frac{d^2\varphi(x)/dx^2}{\big(d\varphi(x)/dx\big)^2},
\end{equation}
holds. The full derivation of Eq. \ref{iterative_g} is included in Appendix \ref{appendix:gradient}. This recursive procedure can be used to approximate $g(x)$ along a trajectory in the asymptotic sense, which means that we need a sufficiently large number of iterations to obtain an accurate approximation of $g(x)$ \cite{chandramoorthy-s3}. In practice, we generate a sufficiently long trajectory, compute first and second derivatives of the map evaluated along the trajectory, and apply Eq. \ref{iterative_g}. We arbitrarily set $g(x_{\rm init}) = 0$, to start the recursive procedure, and 
obtain $g(\varphi(x_{\rm init})).$ The recursion is continued
by setting $x = \varphi(x_{\rm{init}}),$ and so on. For a sufficiently large $K$, the true value of $g(\varphi_K(x_{\rm init}))$ is approached, for almost every initial condition $x_{\rm init}$. 

\subsection{Interpretation of the density gradient iterative formula}
\begin{figure}[H]
    \centering
    \includegraphics[width=0.7\textwidth]{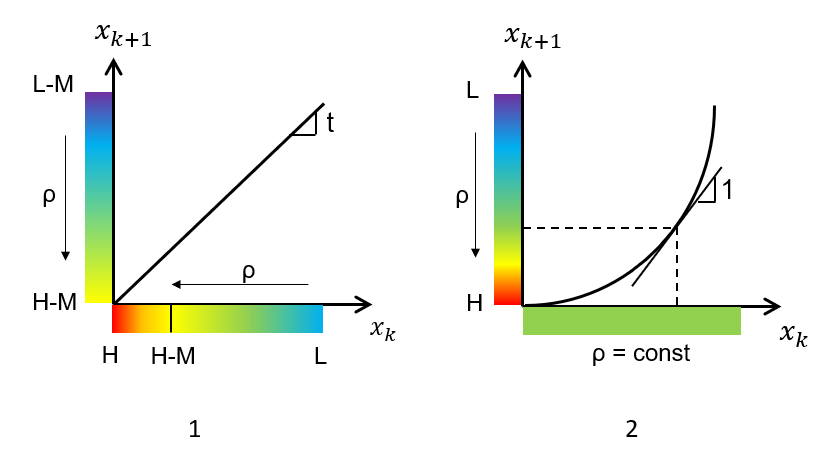}
    \caption{A graphical representation of two different scenarios in one-dimensional maps, 
    to intuitively understand the derivation of the quantity $g$. The bold lines illustrate the map, while shaded bars adjacent to each axis represents the corresponding density distribution on that axis. The region around $H$ corresponds to a high value of density, while the region around $L$ to low values. The slope of the line is indicated as $t$.}
    \label{fig:graphicalg}
\end{figure}

To intuitively understand the density gradient formula (Eq. \ref{iterative_g}), we isolate the effects of each term in Eq. \ref{iterative_g}. In order to do this, we consider a small interval around an iterate $x_k$ and examine two cases: 1. the map is a straight line on this interval and 2. the map has a constant curvature on this interval. These two cases are graphically shown on the left (numbered as 1) and right hand sides (numbered as 2) of Figure \ref{fig:graphicalg}.
%----
%\nisha{we isolate the effects of each term in Eq. \ref{iterative_g}. In order to do this, we consider a small interval around an iterate $x_k$ and examine two cases: 1. the map is a straight line, without curvature, on this interval and 2. the map has a constant curvature on this interval. These two cases are graphically shown on the left (numbered as 1) and right hand sides (numbered as 2) of Figure \ref{fig:graphicalg}.} let us consider two different cases, a linear (left hand side plot, numbered 1)  and nonlinear map (right hand side plot, numbered 2), as depicted in Figure \ref{fig:graphicalg}.
%----
The $x$-axis represents an interval around an iterate $x_k$ and the y-axis, an interval around $x_{k+1}=\varphi(x_k)$. The density $\rho$, around each interval, is shown adjacent to the axes, as a colormap. The colors reflect the distribution of $\rho$ on a logarithmic scale.
%----
%\nisha{The $x$-axis represents an interval around an iterate $x_k$ and the y-axis, an interval around $\varphi(x_k)$. The density $\rho$, around each interval, is shown adjacent to the axes, as a colormap. The colors correspond to $\log{\rho}$.} 
%----
%----
%\nisha{colorbars, one word?} Adam: Yes!
%----
%----
%\nisha{The relationship Eq. 20 should read as 
%\begin{align*}
%    \rho \circ \varphi(x) = \dfrac{\rho}{|\varphi'|}(x)
%\end{align*}}
%----
\begin{enumerate}
   \item 
Consider a small region of $(x-\epsilon, x+\epsilon)$
where the map $\varphi(x)$ has zero
second derivative, i.e., the first derivative $d\varphi/dx$ is constant. As shown in Figure \ref{fig:graphicalg}(1), let us assume that 
the density on the left side of the region, $\rho(x-\epsilon)$, is higher than
the density on the right side, $\rho(x+\epsilon)$. 
%From Eq. \ref{iterative_g}, we know that the mapped density can be calculated using the following equation, since the second term vanishes. 
Due probability mass conservation, the mapped density can be calculated using the following equation,
%----
%\nisha{From Eq. density gradient, we know that mapped density can be calculated using the following equation, since the second term vanishes}
%----
\begin{equation}
\rho(\varphi(x)) = \frac{\rho(x)}{|d\varphi/dx|}.
\end{equation}
Since we consider case $d\varphi/dx>0$, we drop the absolute value. On this interval where the map is a straight line, this statement says that the density around $\varphi(x)$ is a constant multiple of the density around $x$. On the logarithmic scale, the density around $\varphi(x)$ is 
shifted by a constant, when compared to the density around 
$x$ since,
\begin{equation}\label{5:example}
\log\rho(\varphi(x)) = \log\rho(x) - \log \frac{d\varphi}{dx}.
\end{equation}
%----
%\nisha{This relationship is graphically depicted in  Figure %\ref{fig:graphicalg} (1), 
%where the regions marked $H$ and $L$, corresponding to higher and lower %densities, 
%are shifted to the left.}
%----
This relationship is graphically depicted in  Figure \ref{fig:graphicalg} (1), where the regions marked $H$ and $L$, corresponding to higher and lower densities, are shifted to the left. 
Notice that Eq. \ref{5:example} implies that the difference,
on the logarithmic scale, between the higher and lower
densities on the $y$-axis equals the difference between the higher and
lower densities on the $x$-axis.  
%----
%\nisha{However, the small interval is stretched by a factor of $d\varphi/dx$ under one iteration of the map.} 
%----
However, the small interval is stretched by a factor of $d\varphi/dx$ under one iteration of the map.  
Thus, the derivative of the logarithm of
density decreases by a factor of $d\varphi/dx$.  Mathematically, we
can see this by differentiating both sides of the equation with respect to $x$ (and using that $d\varphi/dx$ is constant), 
\begin{align}
 \left(\frac{1}{\rho}\frac{d\rho}{dx}\right)\bigg|_{\varphi(x)} \frac{d\varphi}{dx}
 = \left(\frac{1}{\rho}\frac{d\rho}{dx}\right)\bigg|_x
\end{align}
From the definition of $g$, this reduces to,
\begin{align}
    g(\varphi(x)) = \frac{g(x)}{d\varphi/dx},
\end{align}
%----
%\nisha{which is confirmed by our formula, Eq. formula for $g$, by setting $d^2\varphi/dx^2 = 0$.} 
%----
which is confirmed by our formula, Eq. \ref{iterative_g}, by setting $d^2\varphi/dx^2 = 0$.

%----
%\nisha{Now in order to isolate the effect of curvature of the map on $g$, we consider a curved map and a constant density region. That is, by definition, $g(x) = 0$ in the interval considered. We now describe that $g\circ \varphi(x)$ becomes non-zero on this interval due to the curvature of the map.} 
%----
\item To isolate the effect of curvature of the map on $g$, we consider a curved map and a constant density region. Thus, by definition, $g(x) = 0$ in the interval considered. We now describe that $g\circ \varphi(x)$ becomes non-zero on this interval due to the curvature of the map. Note that Eq. \ref{5:example} still applies, since it is a restatement of probability mass conservation. This means that $\rho$ is reduced by a factor equal to the slope of the map at every point. This is graphically depicted in Figure \ref{fig:graphicalg}, in which we have assumed that $d\varphi/dx$ is an increasing function that crosses the value $1$ at the point indicated using dashed lines. To the left of this point, the density is therefore increased (shown as $H$) around $\varphi(x)$ and to the right, the density is decreased (shown as $L$), when compared to its uniform value around $x$. Note also the larger the first derivative of the map, the lower the density on the y-axis. Again, by taking the derivative of Eq. \ref{5:example} with respect to $x$, and using the definition of $g$, we obtain
\begin{equation}
\label{eqn:curvatureEffectOng}
g(\varphi(x)) = -\frac{d^2\varphi/dx^{2}}{(d\varphi/dx)^{2}}.
\end{equation}
As mentioned in Case 1, the first derivative $(d\varphi/dx)(x)$, 
gives the factor by which a length $dx$ around $x$ is stretched (or compressed) by $\varphi$. The second derivative gives us the change of this stretching (or compression) as a function of $x$.  Thus, the effect of a non-zero second derivative is felt by the \emph{derivative} of the density, and can again be derived from measure preservation or probability mass conservation. 

%----
%\nisha{this is not true. where the slope is 0, $\rho\circ\varphi$ is not defined.}
%----
%If the map slope is zero, in addition to the above assumption, then $g(\varphi(x))=0$ and the density distribution corresponding to the y-axis is constant as well.

\end{enumerate}
%\item Verify for the saw-tooth map. Choose several parameters.

%\item Verify for the cusp map. Choose several parameters.

%-----------Removing the logistic map from this paper-------------
%\item Show divergence for the logistic map.  Discuss the relation to
%the non-smoothness of its statistics as a function of its parameter.
%----------------------------------------------------------------

\subsection{Numerical examples of density gradients}
\begin{figure}[H]
    \centering
    \includegraphics[width=0.6\textwidth]{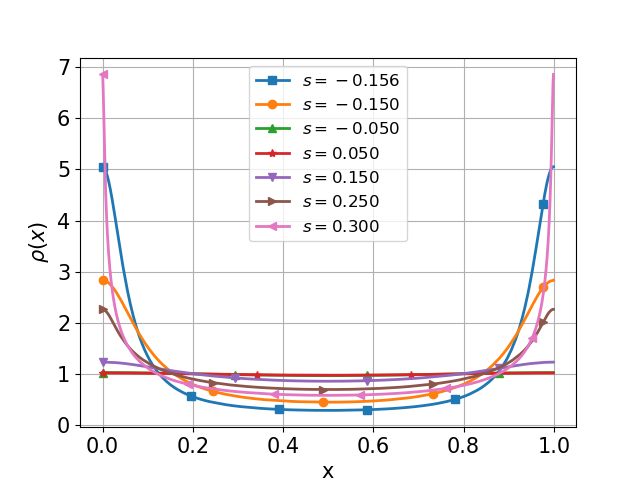}
    \caption{The plot shows the empirically estimated stationary probability distributions achieved by the sawtooth map (Eq. \ref{2:sawtooth}). Every curve was generated using 125,829,120,000 samples and counting the number of solutions in each of 2048 bins of equal length in the interval $[0,1)$.}
    \label{fig:sawtooth_distribution}
\end{figure}

\begin{figure}[H]
    \centering
    \includegraphics[width=0.6\textwidth]{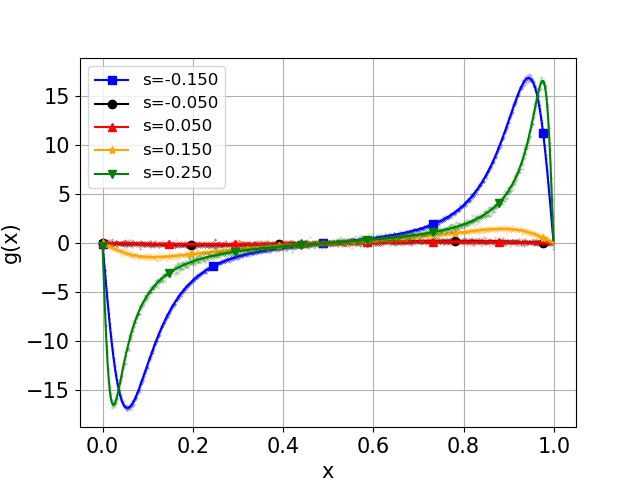}
    \caption{Density gradient function, $g(x)$ (solid lines), generated using Eq. \ref{iterative_g} and compared against the empirically computed value of $g(x)$ (dots), where the derivative of $\rho(x)$ is estimated using finite difference.}
    \label{fig:sawtooth_distribution_gradient}
\end{figure}

In the second part of this section, we show numerical results of the density gradient procedure in two examples, the sawtooth and cusp maps, 
%\nisha{which were introduced in Eq.  and Eq. respectively}
which were introduced in Eq. \ref{2:sawtooth}  and Eq. \ref{eqn:cusp} respectively. Figure \ref{fig:sawtooth_distribution} shows the stationary probability densities of the sawtooth map at different values of $s$. We observe that all curves appear differentiable 
%\nisha{smooth typically means $C^\infty$, so may be change this to ``appear differentiable''?}
, however their derivatives are large, near the interval boundaries, when $s$ is close to $-1/(2\pi)$ or $1/\pi$.
\label{sec:sawtooth}
%\nisha{bins of equal length. Equispaced usually refers to equal spacing between points. spacing between bins here is 0.}
In Figure \ref{fig:sawtooth_distribution_gradient} we show the distribution of the (averaged) density gradient function, $g(x)$, computed using Eq. \ref{iterative_g}, at different values of $s$, and compare it against its finite difference approximation: $(\log(\rho(x+\epsilon)) - \log(\rho(x-\epsilon)))/(2\epsilon).$ Note that the expected value of the density gradient is always zero since
\begin{equation}
    \int_{U} g(x)\;\rho(x)\;dx = 
    \int_{U}\frac{\partial\rho}{\partial x}dx = \left[\rho(x)\right]^{1}_{0}=0.
\end{equation}
%\nisha{in the above caption, ``compared against the empirically computed value of $g$, where the derivative of the empirically computed $\rho$ is estimated using finite difference''}
%----Should we mention heavy tailedness?
%\nisha{becomes significant how? Since the distribution of $g$ is not discussed in this paper, it's probably better to omit this.}
%----

\begin{figure}[H]
    \centering
    \includegraphics[width=0.6\textwidth]{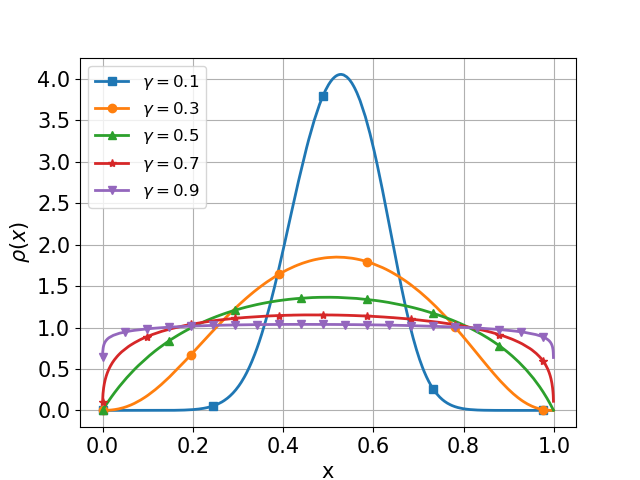}
    \caption{The plot shows the empirically estimated stationary probability distributions achieved by the cusp map (Eq. \ref{eqn:cusp}), at $h=1$ and the indicated value of $\gamma$. All curves were generated in the same fashion as for the sawtooth case.}
    \label{fig:cusp_distribution}
\end{figure}

\begin{figure}[H]
    \centering
    \includegraphics[width=0.6\textwidth]{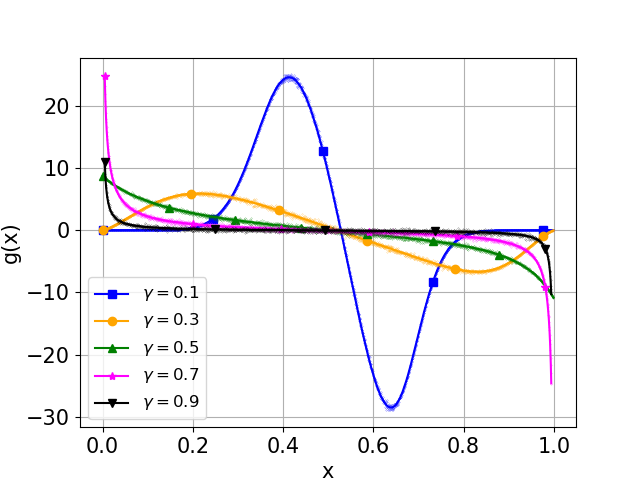}
    \caption{The plot compares $g(x)$ (solid lines) against the derivative of the empirically estimated stationary probability distributions (dots) achieved by the cusp map (Eq. \ref{eqn:cusp}), at $h=1$ and indicated value of $\gamma$. All curves were generated in the same fashion as for the sawtooth case.}
    \label{fig:cusp_distribution_gradient}
\end{figure}

We also repeat a similar experiment for the cusp map, whose results are presented in Figures \ref{fig:cusp_distribution}--\ref{fig:cusp_distribution_gradient}. We observe a behavior similar to the sawtooth map.
%----
%\nisha{Similar to the sawtooth density, the densities computed for the cusp map appear to be differentiable over a range of the parameter $\gamma$. However, as $\gamma$ gets close to 1, the  density $\rho(x)$ acquires large derivatives at the boundaries of the interval. The boundedness of $\rho'(x)$ is needed for the computation of $g(x)$ to be well-conditioned.}
%----
Similar to the sawtooth density, the densities computed for the cusp map appear to be differentiable over a range of the parameter $\gamma$. However, as $\gamma$ gets close to 1, the density $\rho(x)$ acquires large derivatives at the boundaries of the interval. The boundedness of $d\rho/dx$ is needed for the computation of $g(x)$ to be well-conditioned. 
%\subsection{Density gradient of the cusp map}
\label{sec:cusp}

%----------------End of Section 4-----------------------

%---------------------- Section 5-----------------------

\section{Spaces-split sensitivity as a sum of time-correlations}\label{sect:sens}
The evaluation of Eq. (\ref{s3eqn}) is the main focus of this paper.
In practice, expectations with respect to $\rho$, or ensemble averages,
are computed by time-averaging on a single typical trajectory. As mentioned earlier, 
a time average converges to the 
ensemble average of the function, as the length of the trajectory 
approaches infinity. Thus, Eq. \ref{s3eqn} can be written as follows, 
replacing the ensemble averages with ergodic averages
%----
%\nisha{Missing $1/N$ in Eq. 26 and 27}
%----
\begin{align} \label{eqn:s3TimeCorrelation1}
    \frac{d}{ds} \int_U J(x) \,\rho(x)\;dx
  &= -\sum_{k=0}^{\infty} \lim_{N\to\infty} 
  \frac{1}{N}\sum_{n=0}^{N-1} \Big(
    \nabla_{\rho} f(x_n) \, J\big(x_{n+k})\Big),
\end{align}
where $x_i = \varphi_i(x_{\rm{init}})$ is the point at time $i$, along 
a trajectory starting at a typical point $x_{\rm{init}}.$ Using the definition 
of $g$, and taking a long trajectory,
\begin{align} \label{eqn:s3TimeCorrelation2}
    \frac{d}{ds} \int_U J(x) \,\rho(x)\;dx
  &\approx -\frac{1}{N}\sum_{k=0}^{\infty} \sum_{n=0}^{N-1} 
    \Big( \dfrac{df}{dx}(x_n) + f(x_n) g(x_n) \Big)\, J\big(x_{n+k}).
\end{align}
%-----------------------DONE----------------
%\nisha{Digitize Qiqi's graphical explanation figure}
%-------------------------------------------
\subsection{Numerical examples of sensitivities computed using S3}
\begin{figure}[H]
    \centering
    \includegraphics[width=0.6\textwidth]{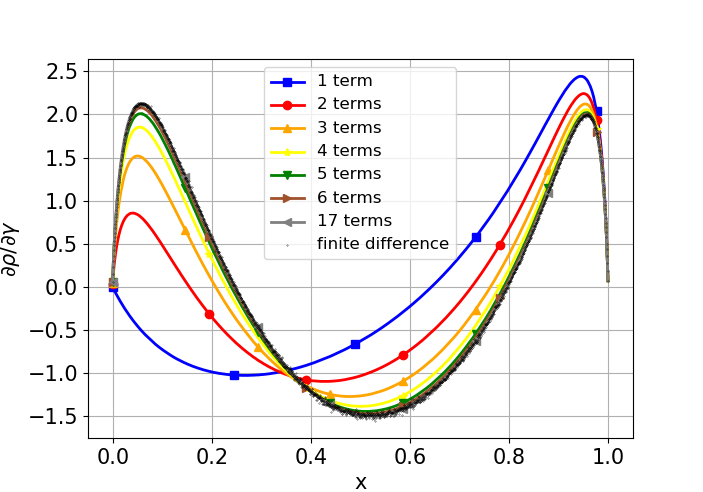}
    \caption{Sensitivity of the density of the cusp map with respect
    to $\gamma$ at $h=1, \gamma=0.5$.  The solid lines represent
    the result of Equation (\ref{s3eqn}) when a finite number of
    terms is used in the summation over $k$.
    The solid line marked with ($\triangleleft$) represents Equation (\ref{s3eqn}) evaluated with 17 terms, which is visibly indistinguishable from the same series summed over 6 or more terms.
    The dots represent
    the finite difference derivative of the density, evaluated
    based on the empirical density at $h=1, \gamma=0.505$
    and at $h=1,\gamma=0.495$.
    Each quantity is evaluated with
    125,829,120,000 samples.}
    \label{fig:cusp_dpower}
\end{figure}

To numerically verify Eq. \ref{eqn:s3TimeCorrelation2}, we consider a set of objective functions, each of which is an 
indicator function denoted by $\delta_c$, and defined such that its value is a constant 1 in a small interval around $c$ and zero everywhere else on the unit interval. With this particular choice, Eq. \ref{eqn:s3TimeCorrelation2} gives us the gradient of the probability density, since 
%\nisha{use align to align the equal signs.}
\begin{equation}
\frac{d}{ds} \int_U \delta_c(x) \,\rho(x;s)\;dx =\int_U \delta_c(x) \frac{\partial\rho(x;s)}{\partial s} \, dx \approx  \frac{\partial\rho(c;s)}{\partial s}.
\end{equation}
Thus, by varying the constant $c$ in the interval $[0, 1)$, 
and using the density gradient computed using Eq. \ref{eqn:iterativeg}, one can compute $d\rho/ds$ over the 
unit interval by using Eq. \ref{eqn:s3TimeCorrelation2}. 
This can be compared with the finite difference approximation
of $d\rho/ds$ generated using slightly perturbed values of 
$s$ and approximating the density empirically. This particular choice of $J(x)$ exhibits yet another advantage of the S3 method over Ruelle's formula (Eq. \ref{ruelle}). The former is also applicable to objective functions that have non-differentiable points, since unlike a direct evaluation of Ruelle's formula, the derivative of $J$ is not used. Figure \ref{fig:cusp_dpower} shows numerical results for the cusp map, in which the density gradient is computed using the space-split formula (Eq. \ref{eqn:s3TimeCorrelation2}) and compared with the central difference derivative.
%\nisha{ and compared with the central difference derivative}. 
We observe that only a few terms of the series are required to produce accurate sensitivities. 
%\nisha{the y-label in Fig 12 must be $\partial \rho/\partial s$}

\begin{figure}[H]
    \centering
    \includegraphics[width=0.6\textwidth]{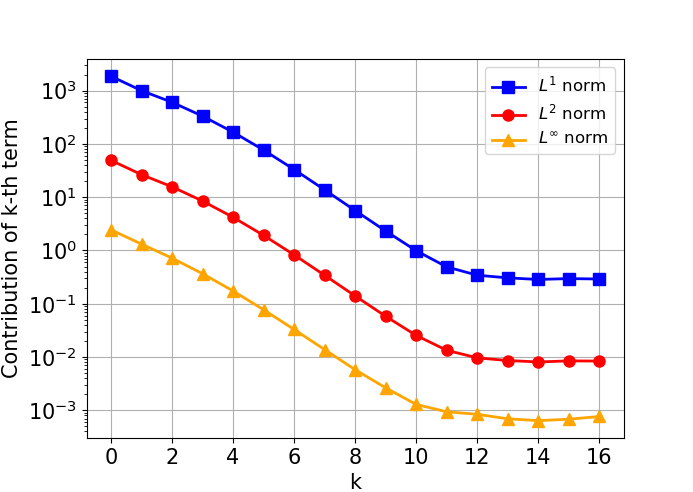}
    \caption{Contributions from the $k$-th term to Equation (\ref{s3eqn}) for the cusp map. Later terms are overwhelmed by
    statistical noise.}
    \label{fig:cusp_dpower_contrib}
\end{figure}
\begin{figure}[H]
    \centering
    \includegraphics[width=0.6\textwidth]{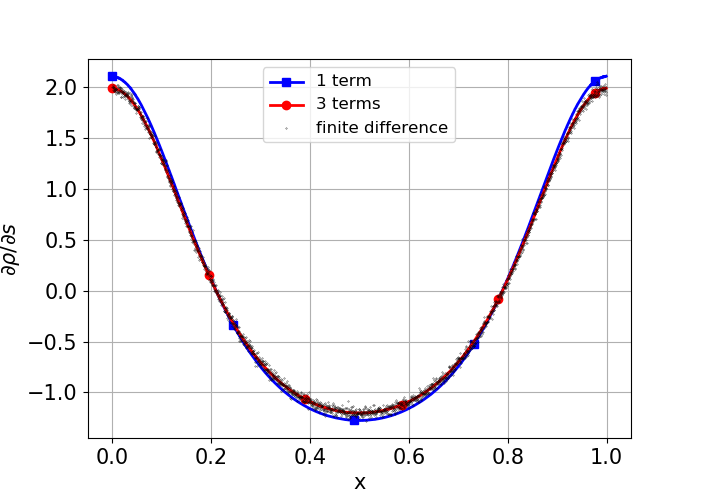}
    \caption{Sensitivity of the density of the sawtooth map with respect
    to $s$ at $s=0.1$.  The solid lines represent
    the result of Eq. \ref{s3eqn} when a finite number of
    terms is used in the summation over $k$.
    The solid line marked with ($\circ$) represent Equation (\ref{s3eqn}) evaluated with 3 terms, is aligned with the the finite difference derivative of the density, evaluated
    based on the density at $s=0.105$
    and at $s=0.095$.
    Each quantity is evaluated with
    125,829,120,000 samples.}
    \label{fig:sawtooth_dpower}
\end{figure}
\begin{figure}[H]
    \centering
    \includegraphics[width=0.6\textwidth]{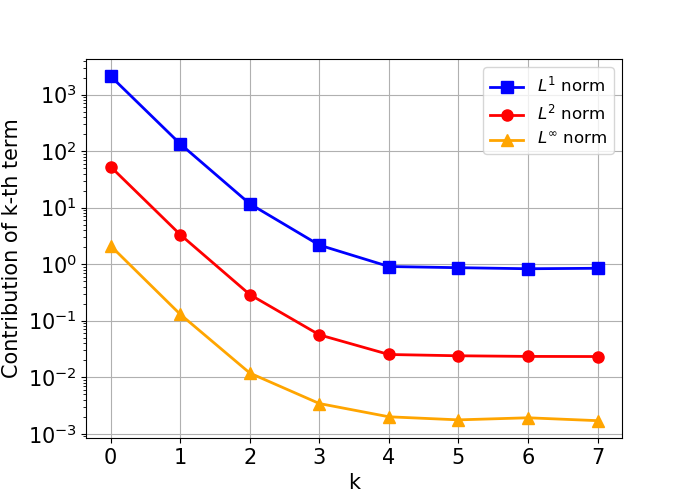}
    \caption{Contributions from the $k$-th term to Eq. \ref{s3eqn}, for the sawtooth map. Later terms are overwhelmed by
    statistical noise.}
    \label{fig:sawtooth_dpower_contrib}
\end{figure}

%----
%\nisha{In the caption of Fig 13, ``contributions from the $k$th term''. Very nice results!}
%----
Figure \ref{fig:cusp_dpower_contrib} clearly indicates that the consecutive terms of the series in Eq. \ref{eqn:s3TimeCorrelation1} exponentially decay in norm. We repeat a similar experiment for the sawtooth map (see Figures \ref{fig:sawtooth_dpower} and \ref{fig:sawtooth_dpower_contrib}). In this case, we only need three terms of Eq. \ref{eqn:s3TimeCorrelation2} to obtain a result that is indistinguishable from its finite difference approximation. The consecutive terms of Eq. \ref{eqn:s3TimeCorrelation2} also decay exponentially in norm.
%----
%\nisha{The y label of Figure 14 must be corrected.}
%----
%----
%\nisha{Contribution from the $k$th term.}
%\nisha{change ``Equation'' in the caption, to Eq. to be consistent with the text.}
%----
%Explain that each term in the summation is the lag-$k$ correlation
%between $\nabla_{\rho}f$ and $J$.
Note that each term of Eq. \ref{s3eqn} is in the form of a 
lag-$k$ time correlation between $\nabla_{\rho}f$ and $J$. We use the term ``lag-$k$'' as $\nabla_{\rho} f$ and the objective function are evaluated at two different states that are k steps apart. In mixing systems, the lag-k time correlations converges to zero as $k\to\infty$. Moreover, for a family of dynamical systems known as Axiom A, the rate of decay of time correlations is proven to be exponential \cite{young-srb}.  In the case of one-dimensional maps, Axiom A systems are the ones in which the derivative of the map is different than 1 everywhere. All the map examples we consider in this paper satisfy this requirement. This guarantees that only a small number of time correlation terms are needed to secure high accuracy of the sensitivity approximation.

\subsection{Computational performance of S3}
\begin{figure}[H]
    \centering
    \includegraphics[width=0.6\textwidth]{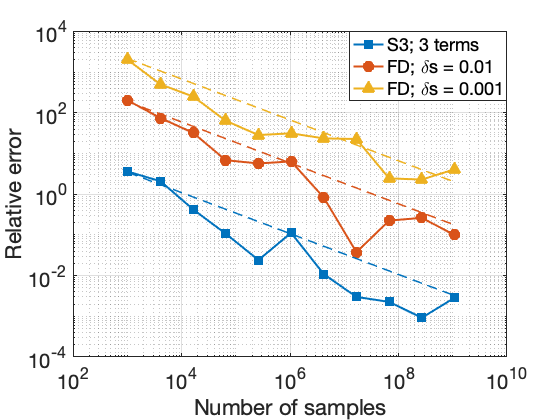}
    \caption{Relative error of the space-split and finite difference methods as a function of the trajectory length. We compute the parametric derivative of density of the sawtooth map at $s=0.1$ on the left boundary ($x=0$). For the S3 computation (curve marked with ($\square$)), we consider only first three terms of Eq. \ref{s3eqn}, which corresponds to the line marked with ($\circ$) in Figure \ref{fig:sawtooth_dpower}. For the finite difference approximation, we calculate densities at $s=0.105, 0.095$ (curve marked with ($\circ$)) and $s=0.1005, 0.0995$ (curve marked with ($\bigtriangleup$)). We also computed the S3 approximation using 125,829,120,000 samples and 9 terms of Eq. \ref{s3eqn}, which serves as a reference value. The dashed lines are proportional to the inverse of the square root of the number of samples.} 
    \label{fig:s3convergence}
\end{figure}

\begin{figure}[H]
    \centering
    \includegraphics[width=0.6\textwidth]{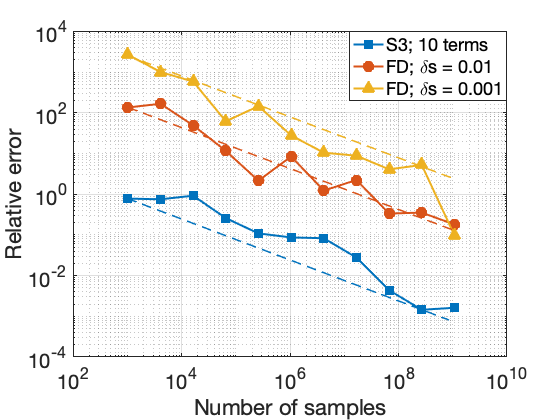}
    \caption{Relative error of the space-split and finite difference methods as a function of the trajectory length. We compute the parametric derivative of density of the cusp map at $h=1$, $\gamma=0.5$ in the middle of the domain $U$ ($x=0.5$). For the S3 computation (curve marked with ($\square$)), we consider first ten terms of Eq. \ref{s3eqn}. For the finite difference approximation, we calculate densities at $\gamma=0.505, 0.495$ (curve marked with ($\bigcirc$)) and $\gamma=0.5005, 0.4995$ (curve marked with ($\bigtriangleup$)). We also computed the S3 approximation using 125,829,120,000 samples and 17 terms of Eq. \ref{s3eqn}, which serves as a reference value. The dashed lines are proportional to the inverse of the square root of the number of samples.} 
    \label{fig:s3convergence_cusp}
\end{figure}

Finally, we compare the space-split sensitivity and classical finite difference method in terms of computational efficiency. We observe in Figures \ref{fig:s3convergence}--\ref{fig:s3convergence_cusp}, generated for the sawtooth and cusp map, respectively, that the S3 method clearly outperforms its competitor, as it requires a few orders of magnitude lesser samples to generate a result with a similar relative error. This is a very promising observation in the context of analysing higher-dimensional systems, 
%\nisha{higher dimensional}
since the large cost of generating very long trajectories
can make such computations infeasible.
%---Removing discussion of computational complexity----
%scales quadratically with the dimension of a map \footnote{The total cost of generating a trajectory is proportional to $Nd^{2}$, where $N$ denotes the trajectory length, while $d$ is the system dimension.}.
%------------------------------------------------------
%\nisha{no, not exponentially.}.
Note in the case of both the S3 and finite difference methods, the error is upper-bounded as follows \cite{chandramoorthy-s3},
\begin{equation}
\mathrm{error}\leq \frac{C}{\sqrt{N}},
\end{equation}
where $N$ denotes the number of samples, while C is some positive number. This means we observe a 
%\nisha{convergence rate of a typical Monte Carlo simulation}
convergence rate of a typical Monte Carlo simulation in both methods. However, the factor $C$ is substantially larger in case of finite differencing. Moreover decreasing the step size (indicated as $\delta s$) in the finite difference calculation, worsens the accuracy, due the dominance of statistical noise.

%\nisha{Nice result!}
%\end{itemize}

%----------------End of Section 5-----------------------

%---------------------- Section 6-----------------------

\section{Conclusions and future work}\label{sect:concl}
We demonstrate a new method to compute the statistical linear response of chaotic systems, to changes in input parameters. This method, known as space-split sensitivity or S3, is used to compute the derivatives with respect to parameters of the long-time average of an objective function. In the S3 method, 
a quantity called \emph{density gradient}, defined as the derivative of the log density with respect to the state, is obtained using a computationally efficient ergodic averaging
scheme. An intuitive explanation of this iterative ergodic averaging scheme, based on probability mass conservation, is discussed in this paper. The density gradient plays a key role in the computation of linear response. Specifically, the sum of 
time correlations between the density gradient and the objective function partially determines the derivative of the mean statistic of the objective function with respect to the parameter. The computational efficiency of the S3 formula when compared to finite difference, which requires several orders of magnitude more samples, stems precisely from this new formula to efficiently estimate the density gradient.

In this work, we restrict ourselves to expanding maps in 1D, which are simple examples of chaotic systems. These examples nevertheless give rich insight into 
chaotic linear response, and specifically into the behavior of the 
density gradient. Our study shows that in same cases the derivative of the density gradient might be very large, which corresponds to heavy tailedness of the density gradient distribution. This phenomenon, as well as its implication for analysis of higher-dimensional maps, is the main topic of our future work.

%----------------End of Section 6-----------------------

%-----Guidelines for the summary------------------------
%\begin{itemize}
%item Summarize ideas and assumptions.
%
%\item Discuss advantages compared to other existing methods

%\item Illustrate additional challenges when applied to %multi-dimensional systems.
%\end{itemize}
%-------------------------------------------------------

\section*{Acknowledgments}
This work was supported by Air Force Office of Scientific Research Grant No. FA8650-19-C-2207.

\bibliographystyle{plain}
\bibliography{refs.bib}

\appendix
\renewcommand{\thesection}{A.\arabic{section}}
%---------------------------Appendix A--------------------------------
\section{Binary floating point problem in simulating 1D maps}\label{appendix:floatingpoint}
\noindent
Consider the case $s=0$. Map \ref{2:sawtooth} can be compactly expressed using the modulo operator, i.e. $x_{n+1}=2x_{n}\hspace{1mm}\mathrm{mod}\hspace{1mm}1$. It means we multiply $x_{n}$ by 2 and if $x_{n+1}>1$, then we also subtract 1. Using floating point arithmetic, we will observe that there exist $N>0$ such that $x_{n}=0$ for all $n\geq N$, which contradicts the assumption of chaotic behavior. This phenomenon is due to the round-off errors associated with the modulo operator. To circumvent this problem, one can change the divisor parameter (of the modulo operation) from 1 to $1 - \epsilon$, where $\epsilon$ is a small number, e.g. $\epsilon=10^{-6}$. Another possible (and simple) workaround might be a change of variables such that the domain of the new variable has irrational length. Note this approach would also require a modification of the objective function. 
%--------------------------Appendix B---------------------------------
\setcounter{equation}{0}
\section{Derivation of the  S3 formula for 1D maps}\label{appendix:s3rederivation}
\noindent
In this section, we will show Eq. \ref{ruelle}--\ref{xspace} are equivalent to Eq. \ref{s3eqn}--\ref{rhogrelation}. Throughout this derivation we will use a short-hand notation for the composition
%----
%\nisha{$v_n := v\circ\varphi^k$, where $v$ is a scalar function}
%----
$v\circ\varphi_{k}=v_{k}$, where $v$ is some scalar function defined on $U=(0,1)$, while $k$ is some integer. If $k=0$, the subscript is dropped.
%----
%\nisha{on $U$}
%----
First, note
%----
%\nisha{Introduce subscript notation in Appedix B instead of C.}
%----

%----
%\nisha{Use this notation throughout: $$\int_{U}f\; \frac{d J_k}{dx}  \; d\mu$$}
%----
\begin{equation}\label{A:1}
\begin{split}
    \int_{U}f\;\frac{dJ_{k}}{dx}\;\rho\;dx =  \int_{U}\frac{d}{dx}\left(f\;J_{k}\right)\rho\;dx-\int_{U}J_{k}\frac{df}{dx}\;\rho\;dx.
\end{split}
\end{equation}
Integrate the first term of Eq. \ref{A:1} by parts,
\begin{equation}\label{A:2}
\begin{split}
\int_{U}\frac{d}{dx}\left(f\;J_{k}\right)\;\rho\;dx = \left[ f\;J_{k}\;\rho\right]^{U_{R}}_{U_{L}} - \int_{U}f\;J_{k}\frac{\partial\rho}{\partial x}\;dx,
\end{split}
\end{equation}
where $U_{L}=0$ and $U_{R}=1$ correspond to the left and right boundary of $U$, respectively. Since the domain is periodic, the first term of Eq. \ref{A:2} vanishes. Thus, we can combine Eq. \ref{A:1} and Eq. \ref{A:2} to conclude that
\begin{equation}\label{A:3}
\begin{split}
    \int_{U}f\;\frac{dJ_{k}}{dx}\;\rho\;dx = - \int_{U}J_{k}\left(\frac{\partial f}{\partial x}+\frac{1}{\rho}\frac{\partial\rho}{\partial x}\right)\;\rho\;dx.
\end{split}
\end{equation}

\setcounter{equation}{0}
%-----------------------------Appendix C------------------------------
\section{Derivation of the iterative procedure for $g$ in 1D maps}\label{appendix:gradient}
\noindent
The purpose of this section is to derive the iterative procedure to calculate the density gradient $g$. We use the same notational convention as in Appendix \ref{appendix:s3rederivation}. Let us consider a function $h$ that is integrable in $U=(0,1)$ and vanishes at $U_{L}=0$ and $U_{R}=1$. Using the definition $g=(1/\rho)(\partial\rho/\partial x)$, and integrating by parts, we obtain
\begin{equation}\label{C:1}
    \int_{U}g\;h\;\rho\;dx = \int_{U}h\;\frac{d\rho}{d x}\;dx = \left[h\;\rho\right]^{U_{R}}_{U_{L}}-\int_{U}\frac{d h}{d x}\;\rho\;dx = -\int_{U}\frac{d h}{d x}\;\rho\;dx.
\end{equation}
The key property used in this derivation is the density preservation of $\varphi$. We say that the map $\varphi$ is density-preserving with respect to the density $\rho$, if for any scalar observable $f$, $\int_U f \; \rho \; dx = \int_U f\circ\varphi_k \; \rho\; dx$ holds for any integer $k$. This implies the left-hand side of Eq. \ref{C:1} can be expressed as
\begin{equation}\label{C:2}
    \int_{U}g\;h\;\rho\;dx = \int_{U}g_{1}\;h_{1}\;\rho\;dx.
\end{equation}
We now apply the density preservation together with the chain rule to the right-hand side of Eq. \ref{C:1}, which gives rise to
\begin{equation}\label{C:3}
    -\int_{U}\frac{d h}{d x}\;\rho\;dx = -\int_{U}\left(\frac{d h}{d x}\right)_{1}\;\rho\;dx = - \int_{U}\frac{d h_1}{dx}\;\frac{1}{d\varphi/dx}\;\rho\;dx.
\end{equation}
Note 
\begin{equation}\label{C:4}
    \frac{dh_1}{dx}\;\frac{1}{d\varphi/dx} = \frac{d}{dx}\left(\frac{h_{1}}{d\varphi/dx}\right) -
    h_{1}\frac{d}{dx}\left(\frac{1}{d\varphi/dx}\right) =
    \frac{d}{dx}\left(\frac{h_{1}}{d\varphi/dx}\right) -
    h_{1}\frac{d^{2}\varphi/dx^2}{(d\varphi/dx)^{2}},
\end{equation}
and, using $h_{1}(U_{L})=h_{1}(U_{R})=0$, integrate by parts to get,
\begin{equation}\label{C:5}
-\int_{U}\frac{d}{dx}\left(\frac{h_{1}}{d\varphi/dx}\right)\;\rho\;dx = -\left[\frac{h_{1}}{d\varphi/dx}\;\rho\right]^{U_{R}}_{U_{L}}
+\int_{U}\frac{h_{1}}{d\varphi/dx}\frac{d\rho}{dx}\;dx = 
\int_{U}\frac{h_{1}}{d\varphi/dx}\frac{d\rho}{dx}\;dx.
\end{equation}
Combine Eq. \ref{C:3}--\ref{C:5} to observe that
\begin{equation}\label{C:6}
    -\int_{U}\frac{d h}{d x}\;\rho\;dx = \int_{U}h_{1}\;\left(\frac{g}{d\varphi/dx}-\frac{d^2\varphi/dx^2}{(d\varphi/dx)^2}\right)\;\rho\;dx.
\end{equation}
Finally, by combining Eq. \ref{C:1},\ref{C:2}, and \ref{C:6}, we obtain the following identity,
\begin{equation}\label{C:7}
    \int_{U}h_{1}\;g_{1}\;\rho\;dx = \int_{U}h_{1}\;\left(\frac{g}{d\varphi/dx}-\frac{d^2\varphi/dx^2}{(d\varphi/dx)^2}\right)\;\rho\;dx,
\end{equation}
from which we infer that
\begin{equation}\label{C:8}
    g_{1} = \frac{g}{d\varphi/dx}-\frac{d^2\varphi/dx^2}{(d\varphi/dx)^2}.
\end{equation}

\end{document}